\documentclass[twocolumn]{aastex631}

\usepackage{times}
\usepackage{amsmath,amssymb,amstext}
\hypersetup{linkcolor=red,citecolor=blue,filecolor=cyan,urlcolor=blue}
\usepackage{tabularx}
\usepackage{longtable}
\usepackage{float}
\usepackage{natbib}
\usepackage{graphicx,color}
\usepackage{txfonts}

\usepackage{multirow}
\usepackage{array}
\usepackage{rotating}
\usepackage{sidecap}
\usepackage{hyperref}
\usepackage{epstopdf}
\usepackage{footnote}
\usepackage{tabularx}
\usepackage{booktabs}
\usepackage{longtable}
\usepackage{tabu}
\usepackage{longtable}

\newcommand{\kms}{km\,s$^{-1}$}

\begin{document}

\title{{\large{\bf Kink Oscillation of a Flux Rope During a Failed Solar Eruption}}}
\author{Pankaj Kumar\altaffiliation{1,2}}

\affiliation{Department of Physics, American University, Washington, DC 20016, USA}
\affiliation{Heliophysics Science Division, NASA Goddard Space Flight Center, Greenbelt, MD, 20771, USA}

\author{Valery M. Nakariakov\altaffiliation{1,2}}
\affiliation{Centre for Fusion, Space and Astrophysics, Department of Physics, University of Warwick, Coventry CV4 7AL, UK}
\affiliation{Centro de Investigacion en Astronomía, Universidad Bernardo O'Higgins, Avenida Viel 1497, Santiago, Chile}

\author{Judith T.\ Karpen}
\affiliation{Heliophysics Science Division, NASA Goddard Space Flight Center, Greenbelt, MD, 20771, USA}

\author{C.\ Richard DeVore}
\affiliation{Heliophysics Science Division, NASA Goddard Space Flight Center, Greenbelt, MD, 20771, USA}

\author{Kyung-Suk Cho}
\affiliation{Korea Astronomy and Space Science Institute, Daejeon, 305-348, Korea}
\affiliation{University of Science and Technology, Daejeon 305-348, Korea}

\email{pankaj.kumar@nasa.gov}

\begin{abstract}
 We report a decaying kink oscillation of a flux rope during a confined eruptive flare, observed off the solar limb by SDO/AIA, that lacked a detectable white-light coronal mass ejection. 
The erupting flux rope underwent kinking, rotation, and apparent leg--leg interaction during the event. The oscillations were observed simultaneously in multiple AIA channels at 304, 171, and 193~\AA, indicating that multithermal plasma was entrained in the rope. After reaching the overlying loops in the active region, the flux rope exhibited large-amplitude, decaying kink oscillations with an apparent initial amplitude of 30~Mm, period of about 16 min, and decay time of about 17 min. We interpret these oscillations as a fundamental standing kink mode of the flux rope. The oscillation polarization has a clear vertical component, while the departure of the detected waveform from a sinusoidal signal suggests that the oscillation could be circularly or elliptically polarized. The estimated kink speed is 1080 \kms, corresponding to an Alfv\'en speed of about 760 \kms. This speed, together with the estimated electron density in the rope from our DEM analysis, $n_e \approx$~(1.5--2.0)~$\times 10^9$~cm$^{-3}$, yields a magnetic field strength of about 15~G. To the best of our knowledge, decaying kink oscillations of a flux rope with non-horizontal polarization during a confined eruptive flare have not been reported before. These oscillations provide unique opportunities for indirect measurements of the magnetic-field strength in low-coronal flux ropes during failed eruptions. 

\end{abstract}


\section{INTRODUCTION}\label{intro}
Waves and oscillations detected in the solar atmosphere are often discussed in reference to plasma heating and solar-wind acceleration, and they can be used to estimate plasma parameters indirectly by magnetohydrodynamic (MHD) seismology \citep{Roberts2000, DeMoortel2012, 2020ARA&A..58..441N, 2020SSRv..216..140V}. In the solar corona, one of the most studied wave phenomena is kink (transverse) oscillations of coronal loops, which have been detected for decades by space-based instruments \citep{Aschwanden1999, Nakariakov1999}. To date, kink oscillations of solar magnetic structures have been observed in two regimes: large-amplitude rapidly decaying oscillations, and low-amplitude decayless oscillations \citep[e.g.,][]{Nakariakov2021}. The large-amplitude events usually are excited by low coronal eruptions that displace the loops from equilibrium \citep{2015A&A...577A...4Z}. The low-amplitude decayless regime could be driven by random or quasi-steady flows around the loops, and remains a subject of intensive research  \citep[e.g.,][]{2016A&A...591L...5N, 2020A&A...633L...8A, 2020ApJ...897L..35K, 2021MNRAS.501.3017R}. 

In the majority of cases, kink oscillations have been detected to be linearly polarized in the horizontal plane \citep[e.g.,][]{2019ApJS..241...31N}, whereas cases of vertical polarization are rare. On the other hand, vertically polarized kink oscillations have interesting seismological and energy-transport implications \citep{2006A&A...446.1139V, 2006A&A...452..615V}. The polarization directions are defined to be relative to the solar surface. In the context of solar images, therefore, kink oscillations that appear as radial motions in the plane of the sky are vertical, while motions parallel to the solar surface are deemed horizontal.  The first observations of rapidly decaying, vertical kink oscillations of loops, with periods of order 200-400 s, were made by the {\rm Transition Region and Coronal Explorer} \citep{Wang2004,Mrozek2011}. Both events appeared be excited by remote eruptions. The Solar Dynamics Observatory's Atmospheric Imaging Assembly (SDO/AIA;  \citealt{lemen2012}) subsequently detected more instances of vertical kink oscillations in coronal loops excited by fast EUV (shock) waves, also associated with remote flares and coronal mass ejections \citep[e.g.,][]{Aschwanden2011, Kumar2013, White2012, Srivastava2013}. To the best of our knowledge, circularly or elliptically polarized kink oscillations have not been definitively detected yet.

\begin{figure*}
\centering{
\includegraphics[width=16.5cm]{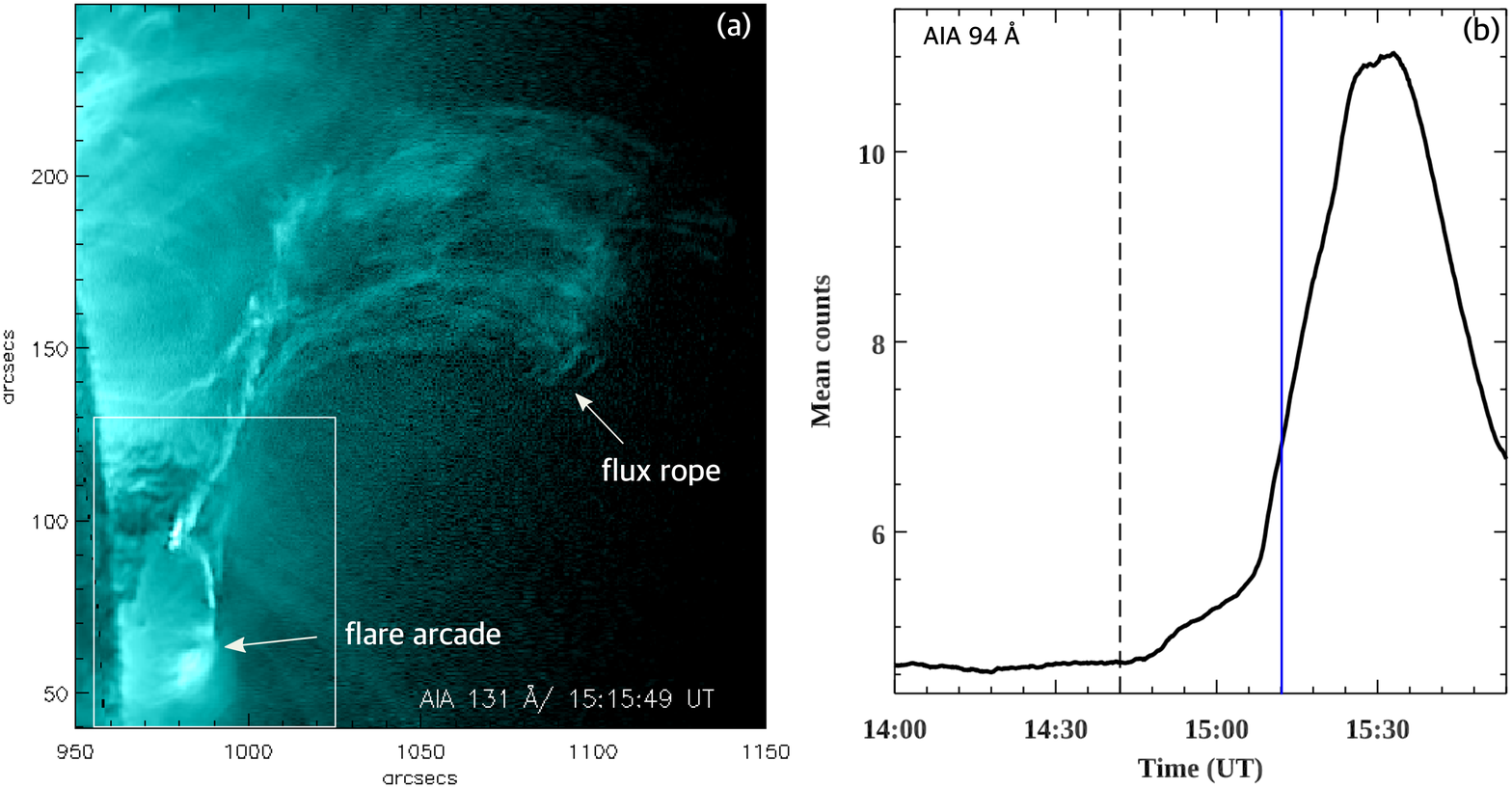}
\includegraphics[width=17cm]{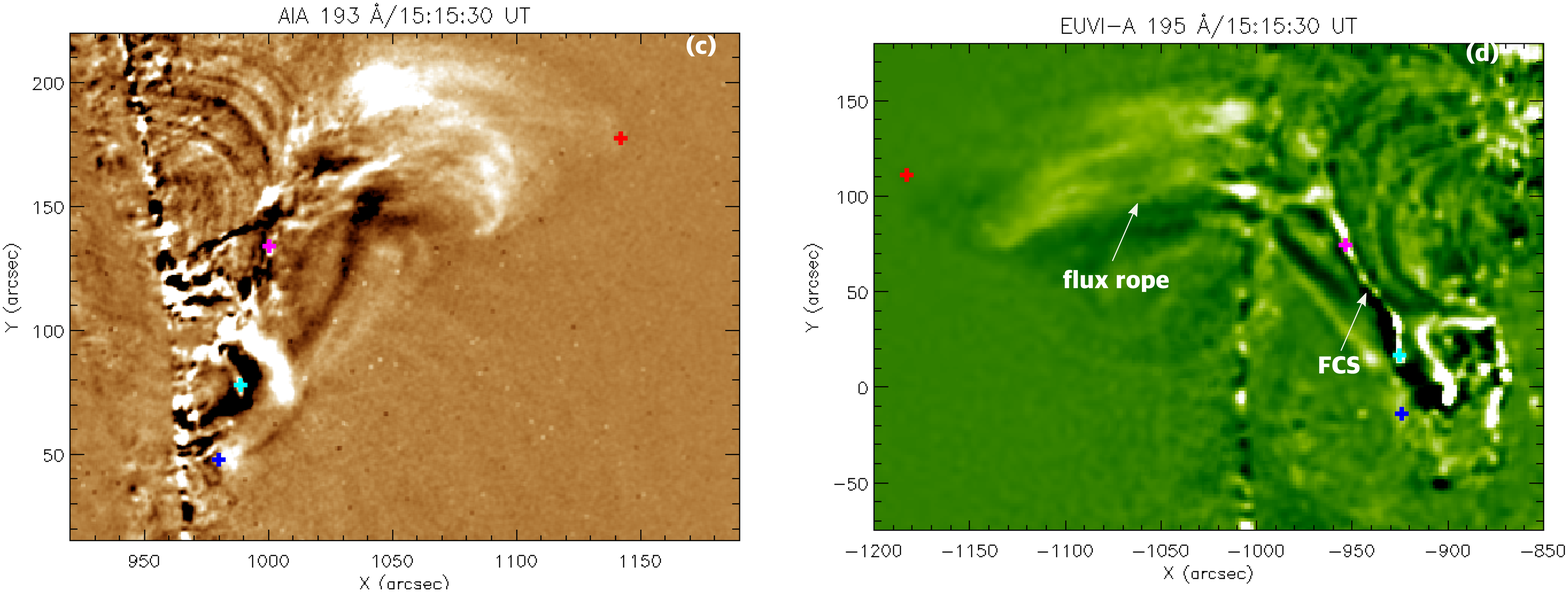}
}
\caption{(a) SDO/AIA~131~{\AA} image showing the rapidly rising flux rope (FR) and associated flare arcade (FA) beneath it. (b) AIA~94~{\AA} mean counts (arbitrary units) extracted from the rectangular box shown in panel (a). The dashed vertical line at 14:42~UT indicates the flare onset, and the solid blue vertical line marks the oscillation start time. (c,d) Cotemporal AIA~193~{\AA} and STEREO/EUVI-A 195~{\AA} running-difference ($\Delta$t=10 min) images. The `+' symbols of different colors indicate the same features in the two images. FCS denotes the flare current sheet. (An animation of AIA 131 \AA~ is available.)} 
\label{fig1}
\end{figure*}

In contrast to kink oscillations of coronal loops, kink oscillations in coronal filaments remain very rarely detected.  \citet{Isobe2006} observed a horizontally polarized kink oscillation of a polar crown filament with a 15~Mm amplitude and 2 hr period during the pre-eruption phase, without any associated flare. The oscillation was possibly excited by magnetic reconnection between a filament barb and nearby emerging flux as seen in {\rm Solar and Heliophysical Observatory}/{\rm Michelson Doppler Imager} magnetograms \citep{Isobe2007}. \citet{2011A&A...531A..53H} observed two successive trains of transverse oscillations with about 100~min period and a similar damping time, in an arched prominence. These oscillations were clearly excited externally, by X-class and C-class flares occurring remotely. \citet{Kim2014} detected a decayless low-amplitude vertical kink oscillation of a flux rope during a failed eruption, which appeared to be excited internally. The oscillation amplitude was comparable to that of decayless kink oscillations of coronal loops \citep{Wang2012, Nistico2013, 2015A&A...583A.136A}. 

Decaying large-amplitude kink oscillations of a coronal magnetic flux rope, associated with the evolution of the rope itself rather than external triggers, have not been reported previously. This phenomenon is of great interest, however, as it provides an excellent opportunity to understand the excitation and decay mechanisms of the oscillation. In addition, measurements of such an oscillation allow estimates of the Alfv\`en speed, magnetic-field strength, and possibly the magnetic twist and electric current in the rope. The lack of direct methods for measuring coronal magnetic fields provides further motivation for identifying and analyzing the rare instances of kink oscillations in a range of solar features. 

Here we report the first observation of a decaying large-amplitude kink oscillation of a flux rope with non-horizontal polarization during a confined eruptive flare. We have analyzed the physical properties and the trigger mechanism of the oscillation. In \S 2, we present the observations and analysis, in \S 3 we discuss the physical interpretation of the oscillating threads, and in \S 4 we summarize the results.


\begin{figure*}
\centering{
\includegraphics[width=17cm]{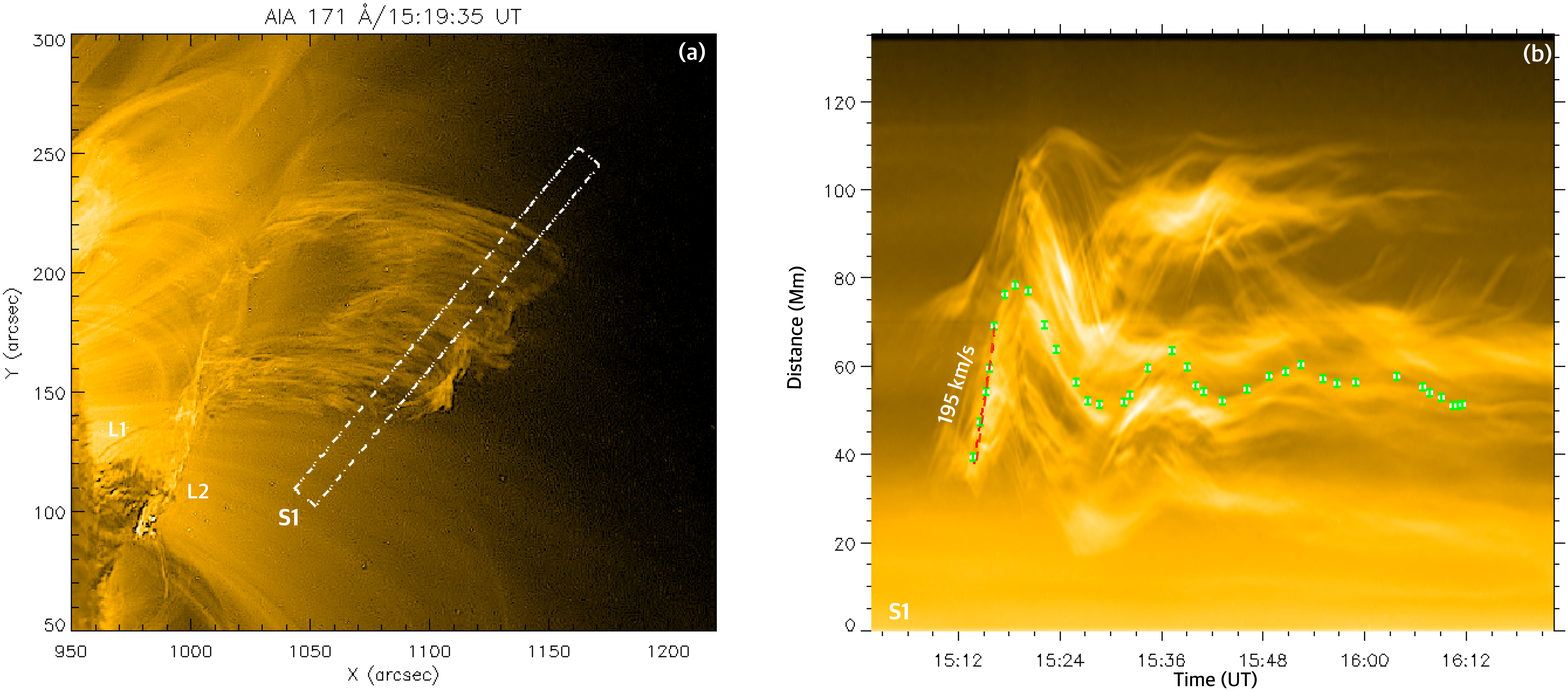}
\includegraphics[width=17cm]{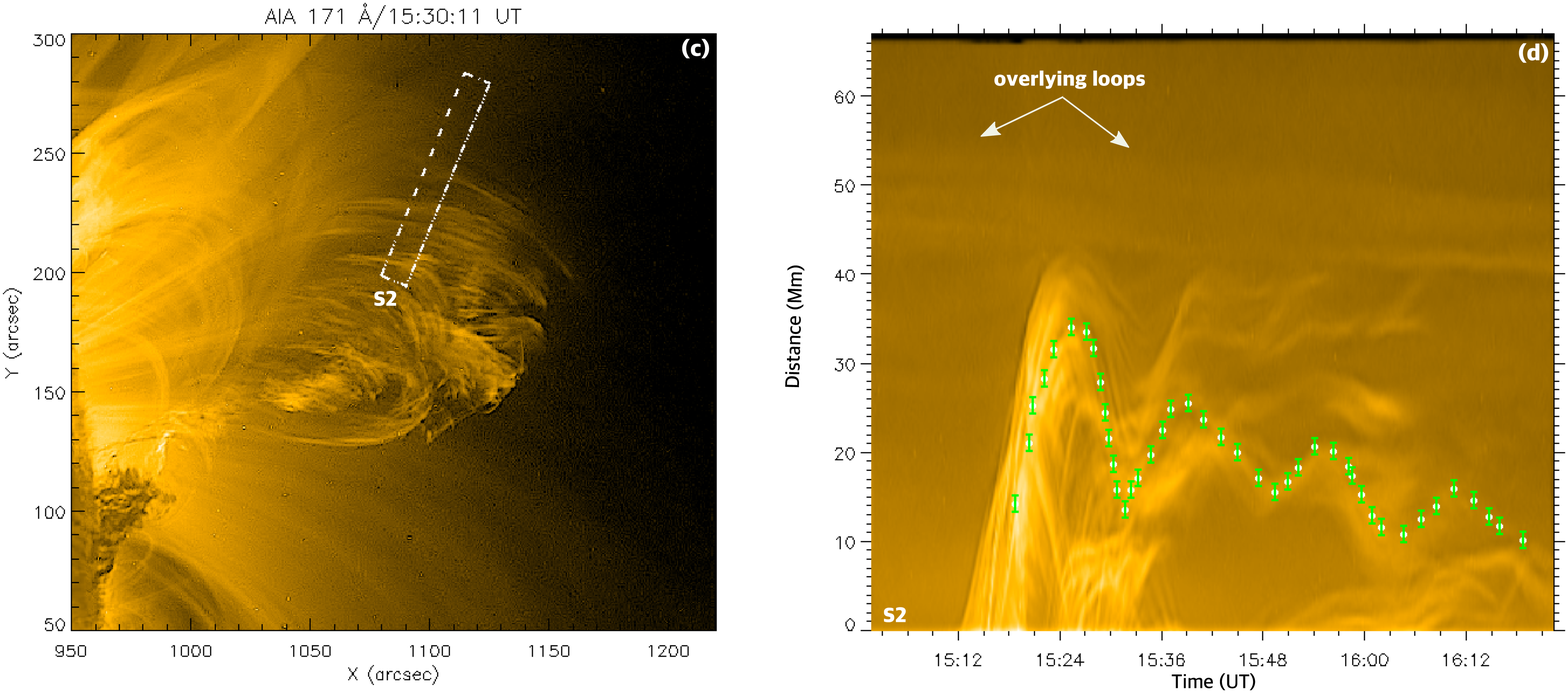}
}
\caption{(a,c) SDO/AIA 171~{\AA} images at two times during the failed eruption, showing the locations of slits S1 and S2 used to construct the time-distance (TD) intensity plots in panels (b, d), which display the kink oscillations of the FR. Arrows in (d) indicate the overlying loops. In panel (a), L1 and L2 mark the two legs of the flux rope. } 
\label{fig2}
\end{figure*}
\begin{figure*}
\centering{
\includegraphics[width=17cm]{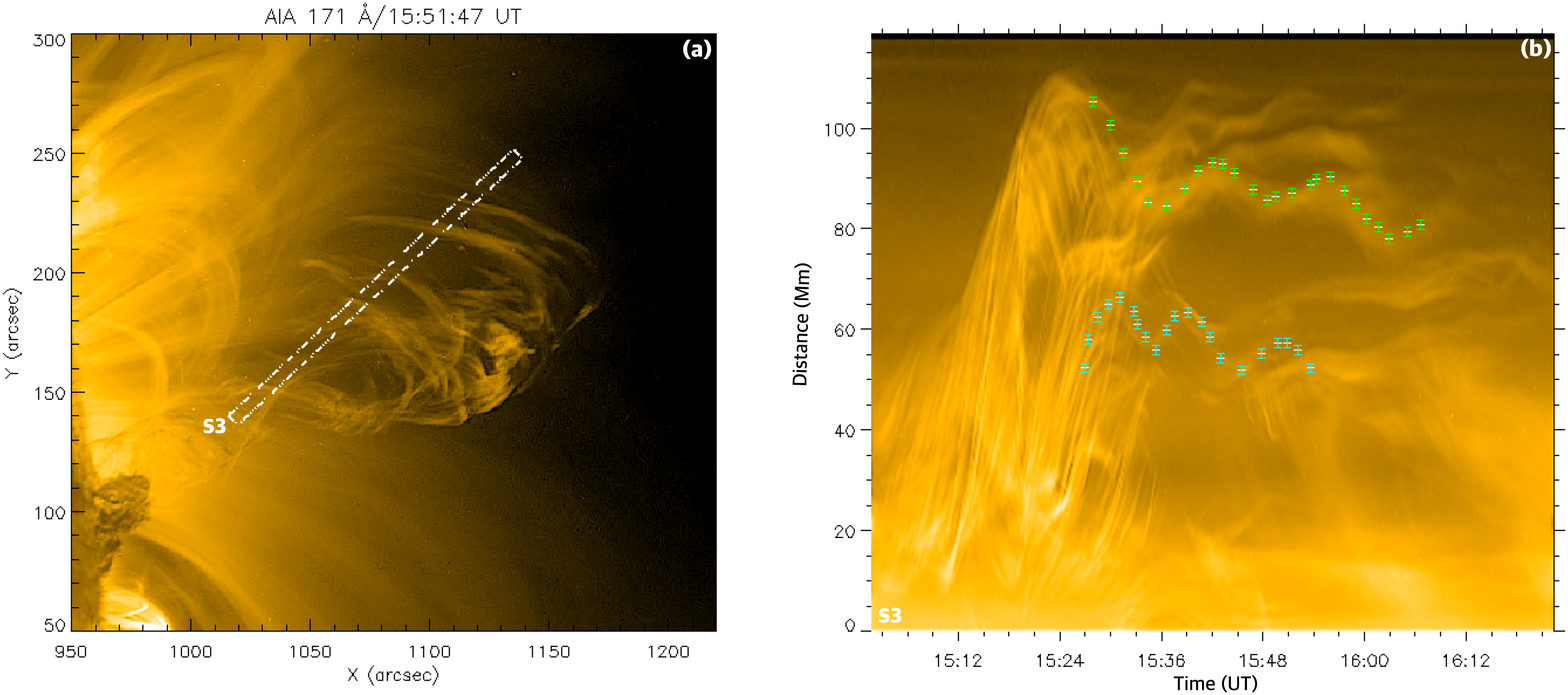}
\includegraphics[width=17cm]{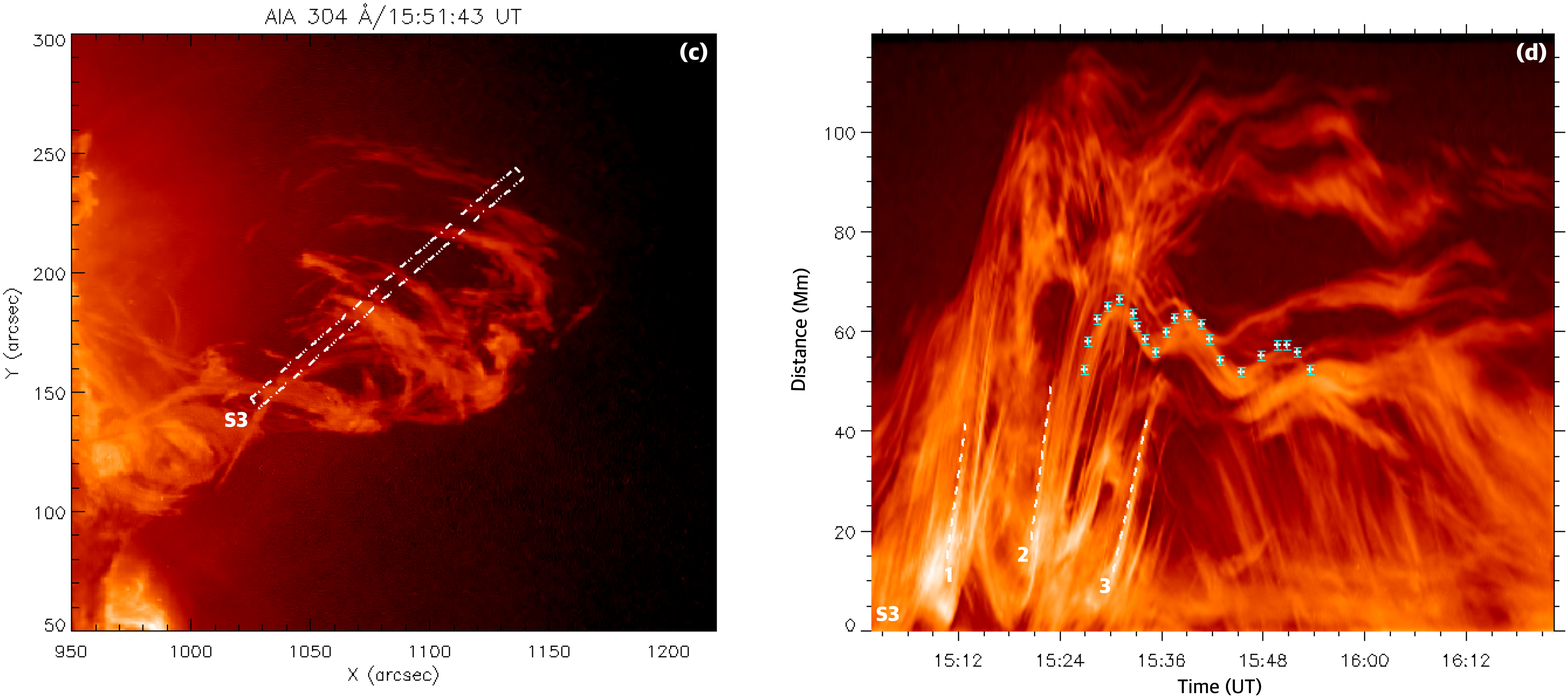}
}
\caption{(a,c) AIA 171 and 304~{\AA} images at the same time during the failed eruption, showing the location of slit S3 used to construct the time--distance (TD) intensity plots in panels (b,d). The green (cyan) symbols highlight the long- (short-)period oscillation. In (d), the dashed lines labeled 1, 2, and 3 indicate outflows originating in the FCS below the rising FR. (An animation of this Figure is available.)} 
\label{fig3}
\end{figure*}

\begin{figure}
\centering{
\includegraphics[width=9cm]{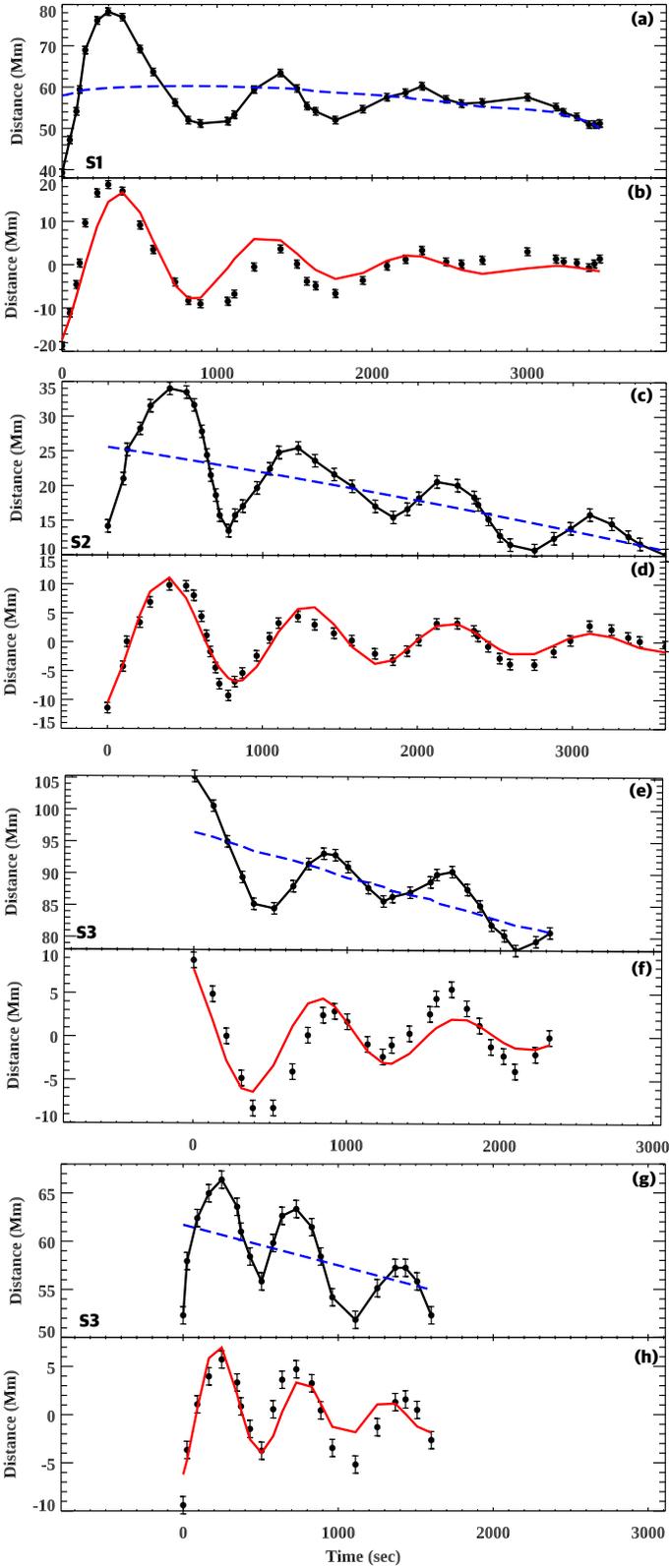}
}
\caption{(a,c,e,g) Waveforms extracted from the TD plots in Figures 2 and 3 (slits S1 to S3). The blue dashed curves show the parabolic trends subtracted to produce the detrended waveforms in panels (b,d,f,h). The blue and red curves are best-fit exponentially decaying sine functions. The start time for each panel is 15:13:49~UT.} 
\label{fig4}
\end{figure}

\begin{figure*}
\centering{
\includegraphics[width=18cm]{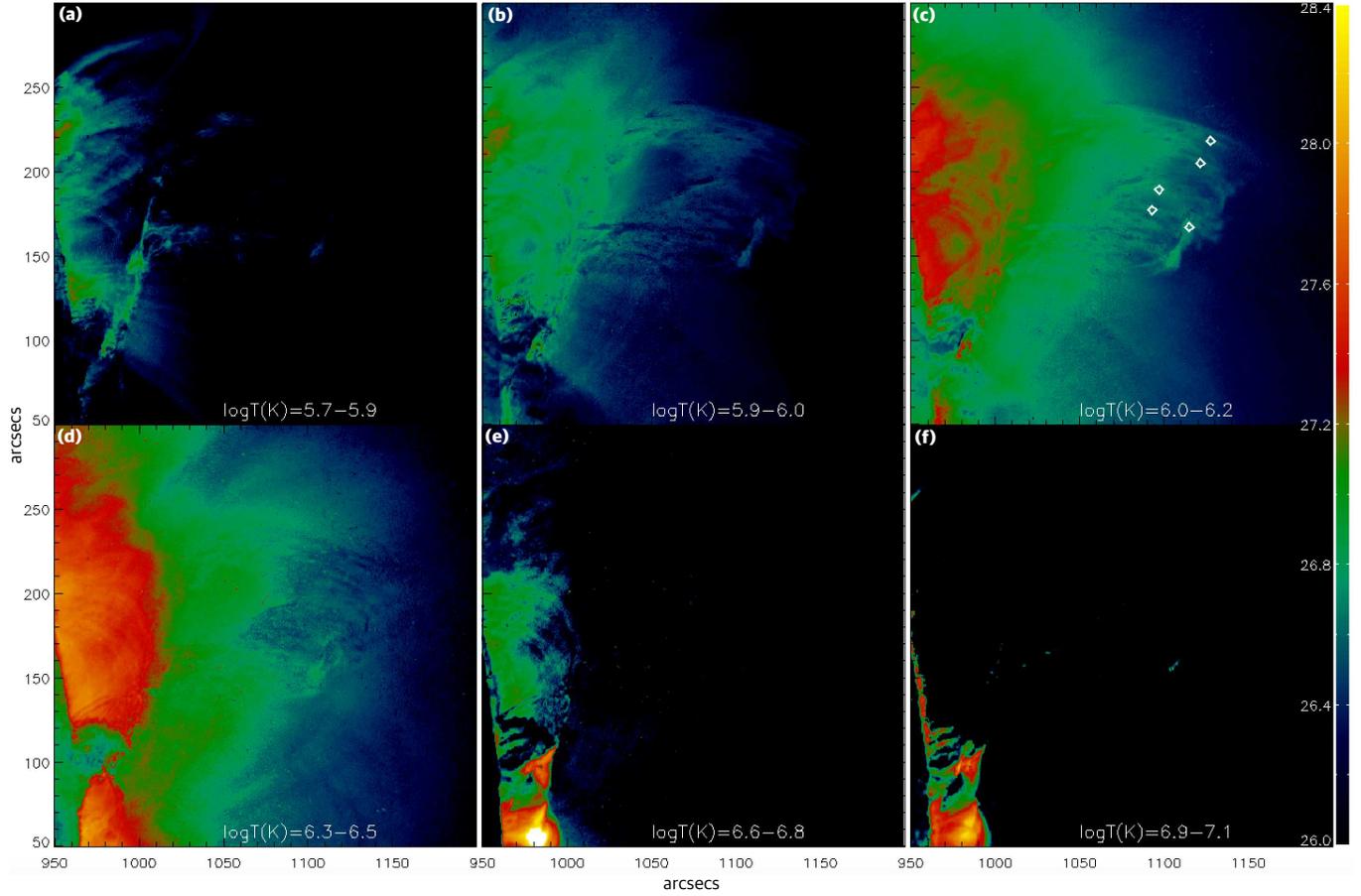}
}
\caption{DEM maps of the FR in different temperature bins using six nearly simultaneous AIA images around 15:19:37~UT. The field of view is the same as in Figures \ref{fig2} and \ref{fig3}.}
\label{fig5}
\end{figure*}
\begin{figure}
\centering{
\includegraphics[width=9cm]{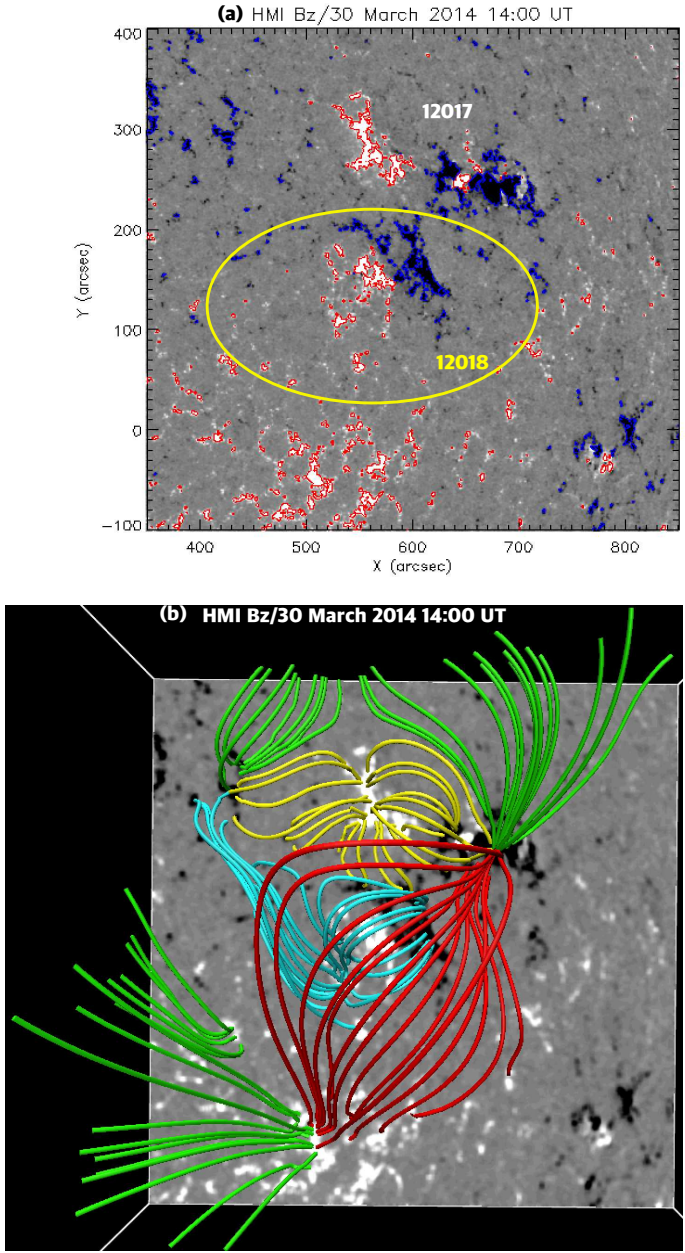}

}
\caption{(a) SDO/HMI magnetogram showing ARs 12017 and 12018 on 2014 March 30. Red/blue shading indicate positive/negative polarities (contour levels = $\pm$100 G). Filament eruption occurred in AR 12018 (marked by yellow ellipse). (b) Potential-field extrapolation of the ARs (same field of view as shown in panel a). Red field lines are arcades overlying the erupting AR 12018 (cyan). Open field lines are shown in green. Yellow field lines represent connecting loops in the neighbouring AR 12017.}
\label{fig6}
\end{figure}

\begin{figure}
\centering{
\includegraphics[width=9cm]{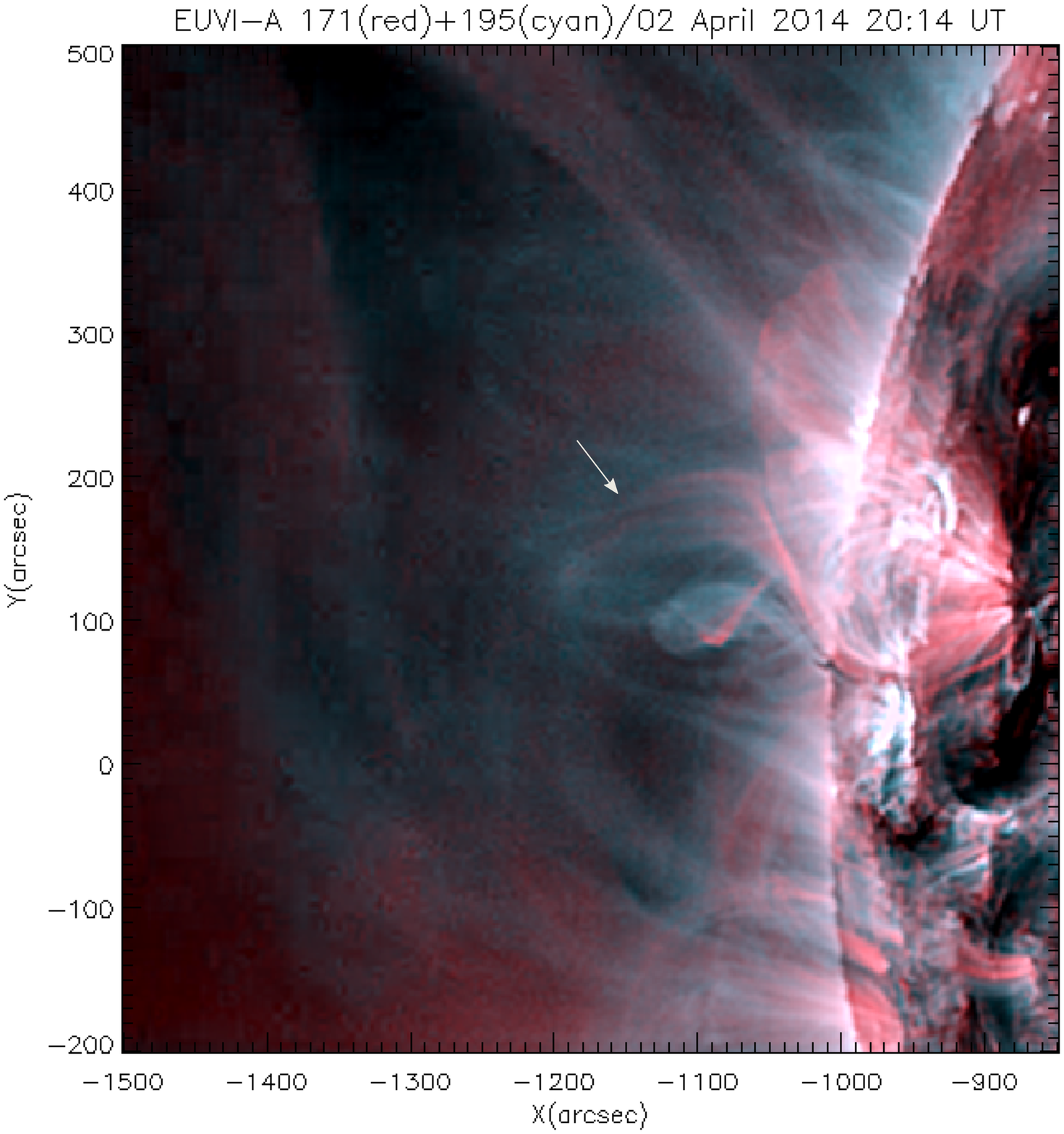}
}
\caption{STEREO/EUVI-A composite image (171 and 195 \AA) on 2014 April 2, one day before the eruption, demonstrating bright overlying loops (marked by an arrow) above an erupting structure from the same filament channel.}
\label{fig7}
\end{figure}


\section{OBSERVATIONS AND ANALYSIS}\label{obs}
We analyzed SDO/AIA full-disk images of the Sun (field-of-view $\approx$ 1.3~$R_\sun$) with a spatial resolution of 1.5$\arcsec$ (0.6$\arcsec$~pixel$^{-1}$) and a cadence of 12~s, in the following channels: 304~\AA\ (\ion{He}{2}, temperature $T\approx 0.05$~MK), 171~\AA\ (\ion{Fe}{9}, $T\approx 0.7$~MK), 193~\AA\ (\ion{Fe}{12}, \ion{Fe}{24}, $T\approx  1.2$~MK and $\approx 20$~MK), and 131~\AA\ (\ion{Fe}{8}, \ion{Fe}{21}, \ion{Fe}{23} at $T\approx 0.4, 10, 16$~MK) images. The 3D noise-gating technique \citep{deforest2017} was used to clean the images. 
To determine the underlying magnetic topology of the investigated region, based on the HMI magnetogram, we utilized a potential-field extrapolation code \citep{nakagawa1972} available in the GX simulator package of SSWIDL \citep{nita2015}. 
The Extreme Ultraviolet Imager \citep[EUVI][]{Wuelser2004,Howard2008} on {\rm Solar TErrestrial RElations Observatory} Ahead (STEREO-A) observed the same event on the disk close to the east limb. The separation angle between SDO and STEREO-A was 155$^{\circ}$ on April 3, 2014. We used EUVI-A 195 {\AA} images (10 min cadence) to observe the magnetic topology of the active region from this different viewing angle. The size of the STEREO/EUVI image is 2048$\times$2048 pixels (1.6$\arcsec$ pixel$^{-1}$), covering a field of view out to 1.7~$R_\sun$. 

The flare and associated failed eruption occurred in active region (AR) NOAA 12018, located at the west limb on 2014 April 3. Figure~\ref{fig1}a displays the eruption and flare underneath it. The GOES soft X-ray (SXR) flux profile in 1-8~{\AA} shows a C-class flare; however, the SXR flux is integrated over the whole solar disk, and there was another flare at the same time from a different AR elsewhere on the disk. Therefore, we use the AIA 94~{\AA} (hot channel) mean counts extracted from a box surrounding the flare arcade (Figure \ref{fig1}a) as a proxy of the flare (Figure \ref{fig1}b), which started around 14:42~UT and peaked at 15:30~UT (Figure~\ref{fig1}b). 
The filament began to rise slowly at about 14:05 UT (see AIA 304~{\AA} movie accompanying Fig.\ \ref{fig3}), well before the flare onset. 
During this slow rise, we do not see evidence of flare reconnection (brightening) underneath it. However, a foreground filament obscures the region closest to the limb, where flare brightenings might have appeared. 
The erupting filament transitioned to fast rise with rotation in its southern leg at 14:42 UT. 
Once the filament reached about 35 Mm above the limb ($\approx$14:43 UT), interchange reconnection began at the breakout current sheet (visible as a bright linear plasma feature) formed from the stressed null where the spine and fan intersect. 
Signatures of this breakout reconnection include bright plasmoids and faint jets.
From 14:55 UT onward, bright blobs traveled outward from the cusp in the bright plasma sheet below the rising structure (see AIA 131 {\AA} movie accompanying Fig.\ \ref{fig1}). 
Bidirectional blobs appeared beneath the rising filament during the flare impulsive phase (15:05-15:25 UT). 
STEREO/EUVI-A images show a rising bubble-like structure above a bright vertical plasma sheet (Fig.\ \ref{fig1}(d)). 
We interpret the bubble-like structure as a flux rope (FR) encompassing the filament, the vertical plasma sheet as a flare current sheet (FCS), and the bright blobs as plasmoids formed during both the breakout and flare reconnection. 
The kink oscillations were detected during the impulsive and decay phases (15:12~UT to 16:12~UT) of the failed eruptive flare.
A detailed analysis of this complex event is in preparation \citep{kumar2022}. 
In the remainder of this paper, we focus on the kink oscillations of the erupting flux rope. 

Using the {\it scc\_measure} (SSWIDL) routine on cotemporal AIA 193 {\AA} and EUVI-A 195 {\AA} running-difference images at 15:15:30 UT, we estimate the true FR maximum height to be 210$\arcsec$ (Figure \ref{fig1}c,d). For comparison, similar structures are marked by plus symbols in both images (Figure \ref{fig1}c,d).  The EUVI-A and AIA observations do not allow us to clearly identify the second leg of the flux rope, due to projection effects and a foreground prominence in the AIA images that obscures the source region. Therefore, we could not accurately measure the distance between the FR footpoints. Assuming a semi-circular shape, the estimated FR length is about 500~Mm (660$\arcsec$).

During the flare impulsive and decay phases (15:12--16:12~UT), the oscillations are best observed in the 304~{\AA} and 171~{\AA} channels (see Figs.\ \ref{fig2} and \ref{fig3} and accompanying movie). The magnetic flux rope rose, bounced back after colliding with overlying loops, oscillated for about 45~min, and ultimately failed to erupt.  Some of the cool and hot plasma fell back toward the surface. Using the movie to determine the locations in which the displacements of the oscillating threads are clearest, we selected slits S1, S2, and S3 for the oscillation analysis (Figs.\ \ref{fig2}, \ref{fig3}). The time--distance (TD) intensity plots were created by averaging the intensity along the width of the rectangular slits. The TD plots along slits S1 to S3 (Figs.\ \ref{fig2}a, \ref{fig2}c, \ref{fig3}a, \ref{fig3}c) reveal decaying oscillations in multiple threads of the FR crossing the slits during the confined eruption. The apparent departure from a sinusoidal waveform seen in S2 in the minimum near 15:30~UT suggests that the oscillation is circularly polarized and viewed with an angle between the line-of-sight and the oscillating loop.  Although the specific kind of polarization cannot be established due to the lack of high-precision observations from different lines of sight, the oscillation clearly is not purely horizontal. The projected initial amplitude of the decaying oscillation is about 30~Mm along S1 (Figure \ref{fig4}a), yielding a projected peak plasma-displacement speed of 195~\kms (red dashed line in Figure \ref{fig2}b). The overlying loop system is visible at 44--54 Mm in S2, marked by arrows in Figure \ref{fig2}d.  During the failed eruption, we also detected multiple outflows along S3 from the plasma sheet below the rising FR. The apparent outflow speeds along the paths marked 1, 2, and 3 in Figure \ref{fig3}d are 204, 125, and 218 \kms, respectively. These outflows are comparable in location and speed to flare reconnection jets observed in many other eruptive events \citep[e.g., ][]{savage2010,takasao2012,kumar2013a,reeves2015,kumar2018,kumar2019b}.

The same oscillation period (within error bars) is seen in the S1 and S2 TDs (Figs.\ \ref{fig2}b,d), but two oscillation periods coexist in the S3 TD  (Fig.\ \ref{fig3}c). The period of the thread situated at 80--100~Mm along S3 is similar to the period seen in the S1 and S2 TDs, but the thread situated at 55--65 Mm in S3 oscillates with a shorter period (highlighted by the white crosses). In this slit, the threads that oscillate with the shorter period seem to be sampled in the legs far from the top of the FR.

We extracted the displacement--time (d-t) oscillatory patterns seen in the TDs (Fig.\ \ref{fig4}a,c,e,g), and subtracted a parabolic trend from those original d-t signals to obtain the detrended d-t oscillation profiles shown in Figure~\ref{fig4}b,d,f,h. The oscillatory profiles were fitted with a decaying sine function to determine the oscillation period and decay time, 
\begin{equation}
d(t)=A\exp\Big(- \frac{t}{\tau}\Big) \sin\Big(\frac{2\pi t }{P}+\phi \Big)+B + Ct,
\label{eq1}
\end{equation}
where $A$, $P$, $\tau$, and $\phi$ are the amplitude, period, decay time, and initial phase, respectively. The best-fit parameters were determined by the least-squared-error method. The best- fitting waveforms are shown by the red curves in Figure \ref{fig4}b,d,f,h. The estimated oscillation periods and decay times are about 16~min and 17~min, respectively, for S1, and 15~min and 26~min for S2 (Table~\ref{table1}). For S3, the oscillation period and damping time of the longer-period oscillation are about 15~min and 21~min, respectively (Fig.\ \ref{fig4}f), while the corresponding shorter-period parameters are 9~min and 14~min (Fig.\ \ref{fig4}h). Interpreting the S3 pair as fundamental and second-harmonic modes of the FR yields a period ratio $P_3/(2P_4) = 15/18 = 0.83$, close to unity. 

We performed a differential emission measure (DEM) analysis \citep{cheung2015} of the erupting flux rope using nearly cotemporal AIA images in six EUV channels (171, 131, 94, 335, 193, 211~\AA) at 15:19:37~UT. The emission measure in different temperature bins is shown in Figure~\ref{fig5}. The multi-thermal FR plasma exhibited temperatures between 0.5--1.6~MK ($\log T \mathrm{(K)}=$5.7--6.2) (Fig.\ \ref{fig5}a,b,c), while the flare arcade (FA) and cusp structure contained hot plasma with $ T \approx$4--10 MK (Fig.\ \ref{fig5}e,f). Note that the DEM code does not utilize the AIA 304 {\AA} channel, and no cooler optically thin lines are observed by SDO. Therefore the estimated electron densities are applicable only to the coronal-temperature plasma in the FR, excluding any unheated filament material.

To estimate the plasma electron density, we selected five threads of the flux rope (regions circled in Fig.\ \ref{fig5}c), and determined the total emission measure by integrating the Gaussian DEM distribution over the entire temperature range for the selected threads. The average value of the total emission measure (EM) is about 8.1$\times$10$^{26}$ cm$^{-5}$ for the five threads in the rope. Taking the average column (thread) width as $w\approx$ 3-5$\arcsec$ (Fig.\ \ref{fig5}c), we estimated the electron density as $n_e=\sqrt{\mathrm{EM}/{w}} \approx$ (1.5-2.0)$\times 10^9$~cm$^{-3}$.

 \section{DISCUSSION}\label{discussion}

Interpreting the oscillations is not straightforward, because no theoretical models to date have found a simple expression for the period of a standing kink wave in a curved, highly twisted flux rope. The fundamental-mode oscillations could be interpreted in terms of the model developed by \citet{2016A&A...590A.120K, 2018JASTP.172...40K}, which links the kink oscillation period with parameters of the magnetic dip under the rope and its mass and electric current. However, this model describes motions near the apex only, and hence is not applicable to higher parallel harmonics located in the FR legs.  Instead, we considered making order-of-magnitude estimates with the standard model, which considers the FR as a low-$\beta$ plasma cylinder embedded in a less dense, low-$\beta$ plasma \citep[e.g.,][]{Roberts2000, Nakariakov2021}. A shortcoming of the cylinder model is the difficulty accounting for the effects of the magnetic twist. To the best of our knowledge, most theoretical studies of the dispersive properties of kink oscillations in plasma cylinders with a twisted field assume that the twist component of the field is much smaller than the parallel component \citep[for recent results, see][]{2020MNRAS.496...67B, 2021RAA....21..126W}. Under these conditions, \citet{2017ApJ...840...26L} found that the kink waves are practically unaffected by the magnetic twist inside the cylinder.

Therefore, we simply link the period with the FR length along the axis between the footpoints $L$ and the kink speed $C_\mathrm{k}$,  
\begin{equation} \label{ck}
P_n \approx 2L/(n C_\mathrm{k}),
\end{equation}
where $n$ is the harmonic number parallel to the curved FR axis. Using  $L\approx 500$~Mm (Figure \ref{fig1}c,d), and the mean period of the fundamental harmonic $P_1 \approx 15$~min, we estimate the kink speed as 1080~\kms. In a low-$\beta$ plasma, the kink speed is determined by the Alfv\'en speed inside the cylinder as $C_\mathrm{k} \approx \sqrt{2}V_\mathrm{A}$ \citep[e.g., ][]{Roberts1983,ofman2008}. With the electron density estimated from the DEM, $n_e \approx$~2.0~$\times 10^9$~cm$^{-3}$, and the implied Alfv\'en speed $V_\mathrm{A} \approx 760$~\kms, the magnetic-field strength in the oscillating flux rope is about 15~G. This value is comparable to previous polarimetric measurements of the average magnetic-field strength in active-region prominences \citep{leroy1989,casini2003, sch2013,mackay2020}.

The oscillations were excited when the flux rope interacted with, and bounced back from, the overlying arcade of loops in the active region. The magnetic reconnection outflows in the vertical flare current sheet were 125--218~\kms (15:05--15:28~UT). These outflows helped to accelerate the flux rope outward, but the overlying arcades of loops did not allow the rope to escape. Therefore, the FR moves up and down in the plane of sky. No EUV wave was detected in association with the eruption, as evidenced by AIA 193/211~{\AA} running-difference images taken during the event. Therefore, we rule out a flare-generated EUV/shock wave \citep{Zimovets2015} as the trigger of the kink oscillations. We infer that the oscillations were excited by the combination of the upward acceleration due to reconnection outflows from the FCS below the FR, and the downward acceleration due to magnetic tension forces in the coronal arcade of loops above the FR.

An intriguing outcome of this study is the decaying behavior of the observed kink oscillations, in contrast to the decayless, low-amplitude, kink oscillations in a FR detected by \citet{Kim2014}. These apparently different regimes could be understood in terms of the self-oscillation model  \citep{2016A&A...591L...5N} confirmed by full-scale MHD simulations  \citep{2020ApJ...897L..35K}. In the decayless regime, a fully saturated self-oscillation is supported by a steady energy supply, for example, by the interaction of the flux rope with an external steady flow or a flow varying on time scales very different from the oscillation period. The decaying-oscillation regime is caused by an impulsive excitation, with the initial amplitude exceeding subsequent peaks, as observed in some coronal loops \citep[e.g., ][]{2013A&A...552A..57N}. Theoretical validation of the response of a self-oscillating magnetic flux rope to an impulsive driver, and detailed comparison with observed events, need to be carried out.

The lack of oscillations in the overlying arcade could be attributed to specific properties of the colliding flux systems. The AIA images only reveal proper motions perpendicular to the line of sight, but there could be a horizontal component as well.  If the FR axis were parallel to the field in the loops, then the rope could push the overlying loops sideways during the collision, and the oscillations excited in the overlying loops would be horizontally polarized along the line of sight.
A more likely option is that the amount of magnetic flux in the overlying arcade was higher than in the rope, making the overlying field harder to perturb. This is consistent with the eruption having failed to escape. In addition, the relative orientation of the field between the top of the flux rope and the overlying arcade is unfavorable for reconnection to remove the overlying field, which would enable a successful eruption. Kinking/rotation generally changes the field direction at the top of the rope. We speculate about the role of the overlying flux because there was a stronger neighboring active region, AR 12017, north of the AR (12018) where the failed filament eruption took place (Fig.\ \ref{fig6}a). The potential-field extrapolation (Fig.\ \ref{fig6}b) of the ARs reveals significant flux (red) overlying the eruption site in AR 12018 (cyan). The northern footpoints of these overlying arcades were connected to the sunspot ($B_{max} \approx -$1200 G) in AR 12017. STEREO images of these ARs (1 day before the eruption, when the active region was on the limb), show many bright loops above the erupting region (Fig.\ \ref{fig7}).  
More observations and simulations of failed eruptions are needed to establish whether this lack of impact on the surrounding field is common, and to determine the polarization of such FR oscillations. 

\section{SUMMARY}\label{summary}

In this study, we detected and analyzed large-amplitude, decaying kink oscillations of a magnetic flux rope in the solar corona. The observed polarization is vertical, circular, or elliptical; the lack of high-resolution observations of the oscillation from different points of view prevents an unequivocal determination of the polarization. Such oscillations of a flux rope with definitely non-horizontal polarization, generated during a failed eruptive flare, have not been reported before. During the observed evolution, the flux rope exhibited kinking, rotation, and evidence for leg-leg reconnection. 
Similar signatures have been reported in MHD simulations of a kink-unstable flux rope \citep{kliem2010}. However, no known MHD simulation to date has shown a decaying kink oscillation of a flux rope during a confined eruptive flare, as presented here. 

The oscillations were detected in multiple AIA channels during the impulsive and decay phases of a flare associated with a failed eruption. The flux rope contained multi-thermal plasma, from cool prominence threads to 10~MK heated coronal material. The length of the rope is $L\approx 500$~Mm, and the electron number density in the oscillating plasma is $n_e \approx$ (1.5-2.0)$\times 10^9$~cm$^{-3}$. The oscillations were excited internally by the collision of the rising flux rope with overlying loops. In the central part of the rope, near the apex, the oscillation period was 15~min and damping time was 21~min. In the lower part of one leg of the rope, we detected a shorter-period oscillation, with period 9~min and damping time 14~min.  Assuming that the detected kink oscillations are collective motions of FR threads with comparable lengths, we interpret them as standing harmonics of the rope, with the longer- and shorter-period oscillations corresponding to the fundamental mode and the second parallel (to the curvilinear axis of the rope) harmonics, respectively \citep[e.g.,][]{2009SSRv..149....3A, Nakariakov2021}. 

More high-resolution observations of additional failed eruptions should reveal how frequently these kink oscillations are triggered in flux ropes that stop in the corona rather than escaping as CMEs, the estimated field strength in the oscillating portions of the flux ropes, and how the oscillations are damped. In the near future, spectral imaging and spectro-polarimetric measurements from the SPICE instrument on Solar Orbiter \citep{muller2013} and coronal magnetic-field observations from the Daniel K.\ Inouye Solar Telescope \citep{tritschler2015} could help to establish the oscillation polarization and magnetic-field strength in oscillating flux ropes. Complementary 3D MHD simulations of failed eruptions would provide deeper understanding of the triggering and the decay of these oscillations. These observational and theoretical advances would shed new light on the conditions and dynamics in such events, and also could lead to greater understanding of why only some eruptive events successfully produce CMEs while others fail.

\begin{acknowledgments}
We thank the referee for insightful comments that
have improved this paper. SDO is a mission for NASA's Living With a Star (LWS) program. Magnetic-field extrapolation was visualized with VAPOR (www.vapor.ucar.edu), a product of the Computational Information Systems Laboratory at the National Center for Atmospheric Research. This research was supported by NASA's Heliophysics Guest Investigator (\#80NSSC20K0265) and GSFC Internal Scientist Funding Model (H-ISFM) programs. V.M.N.\ acknowledges support from the STFC consolidated grant ST/T000252/1.
\end{acknowledgments}

\bibliographystyle{aasjournal}
\bibliography{reference.bib}

{\small

\begin{longtable*}{c c c c c c}
\caption{Properties of the oscillation detected at different slits. } \\
\hline \\
\label{tab1}
Slit   &AIA channel                 &Initial Amplitude                   &Period             &Decay time             \\
        & (\AA)                           &  (Mm)                                   & (Min)              & (Min)                  \\
           
    \hline
 &                                              &                 &                           &                                                                             \\  
S1       &171                        &30                   & $15.6\pm 0.2$                    &$16.8 \pm 1.5$                                                \\    
  
            &                                  &                  &                          &                                                                                   \\
            
S2       &171                        &13                    & $15.2\pm 0.1$                        &$25.8 \pm 3.7$                                                         \\    
           &                               &                  &                              &                                                                                 \\

S3       &171                        & 10                    & $15.0\pm 0.3$                       &$21.0\pm 4.8$                                \\    
           &                               &                  &                              &                                                                                 \\

S3      &304                   &11                      & $9.1 \pm 0.2$                       &$14.3 \pm 3.9$                                                                         \\    
      
        &                                 &                 &                            &                                                                  \\    
 \hline
\label{table1}
\end{longtable*}
}
\end{document}